\documentclass[twocolumn, aps, pra,unsortedaddress]{revtex4-1}

\usepackage{amsmath,graphicx,amssymb}
\usepackage{graphicx} 
\usepackage{xcolor}
\usepackage{soul}

\usepackage{natbib}

\begin{document}

\title{Optical binding with cold atoms}

\author{C. E. M\'aximo} \affiliation{Instituto de F\'{i}sica de S\~ao Carlos, Universidade de S\~ao Paulo, 13560-970 S\~ao Carlos, SP, Brazil}

\author{R. Bachelard} \affiliation{Instituto de F\'{i}sica de S\~ao Carlos, Universidade de S\~ao Paulo, 13560-970 S\~ao Carlos, SP, Brazil}

\author{R. Kaiser} \affiliation{Universit\'e C\^ote d'Azur, CNRS, INPHYNI, 06560 Valbonne, France}

\date{\today}
\begin{abstract}
Optical binding is a form of light-mediated forces between elements of matter which emerge in response to the collective scattering of light. Such phenomenon has been studied mainly in the context of equilibrium stability of dielectric spheres arrays which move amid dissipative media. In this letter, we demonstrate that optically bounded states of a pair of cold atoms can exist, in the absence of non-radiative damping. We study the scaling laws for the unstable-stable phase transition at negative detuning and the unstable-metastable one for positive detuning. In addition, we show that angular momentum can lead to dynamical stabilisation with infinite range scaling.
\end{abstract}
\maketitle

\section{Introduction}
The interaction of light with atoms, from the microscopic to the macroscopic scale, is one of the most fundamental mechanisms in nature. After the advent of the laser, new techniques were developed to manipulate precisely objects of very different sizes with light, ranging from individual atoms~\cite{kimble1994} to macrosopic objects in optical tweezers~\cite{ashkin1970}. It is convenient to distinguish two kinds of optical forces which are of fundamental importance: the radiation pressure force, which pushes the particles in the direction of the light propagation, and the dipole force, which tends to trap them into intensity extrema, as for example in optical lattices. Beyond single-particle physics, multiple scattering of light plays an important role in modifying these forces. For instance, the radiation pressure force is at the origin of an increase of the size of magneto-optical traps~\cite{wieman1990} whereas dipole forces can lead to optomechanical self-structuring in a cold atomic gas~\cite{labeyrie2014} or to optomechanical strain~\cite{davison2017}.

For two or more scatterers, mutual exchange of light results in cooperative optical forces which may induce optical mutual trapping, and eventually correlations in the relative positions of the particles at distances of the order of the optical wavelength. This effect, called optical binding, has been first demonstrated by Golovchenko and coworkers~\cite{golovchenko1989,golovchenko1990}, using two dielectric microsized-spheres interacting with light fields within dissipative fluids. Since then a number of experiments with different geometries and with increasing number of scatterers have been reported, all using a suspension of scatterers in a fluid providing thus a viscous damping of the motion of the scatterers~\cite{vesperinas2001, ritsch-marte2003, jin2006, wright2006, karasek2006, wright2007, pavel2010, fournier2014}.

Equilibrium is reached when viscous friction dominates the light-induced dynamics, a regime on which previous studies have focused. For instance, bistability of equilibrium separations~\cite{wright2006} and one-dimensional optically restoring forces~\cite{wright2007} were observed in a system of two dielectric particles suspended in a viscous fluid. Furthermore, theoretical simulations have predicted stable configurations for microspheres arrays in two-dimensional (2D) compositions~\cite{vesperinas2001,jin2006}. Actually, out of equilibrium bound motion with light-mediated forces has been assumed to be possible only above a critical damping~\cite{sheng2005}. 

\begin{figure}[htb]	
	\begin{center}
		\includegraphics[scale=0.31]{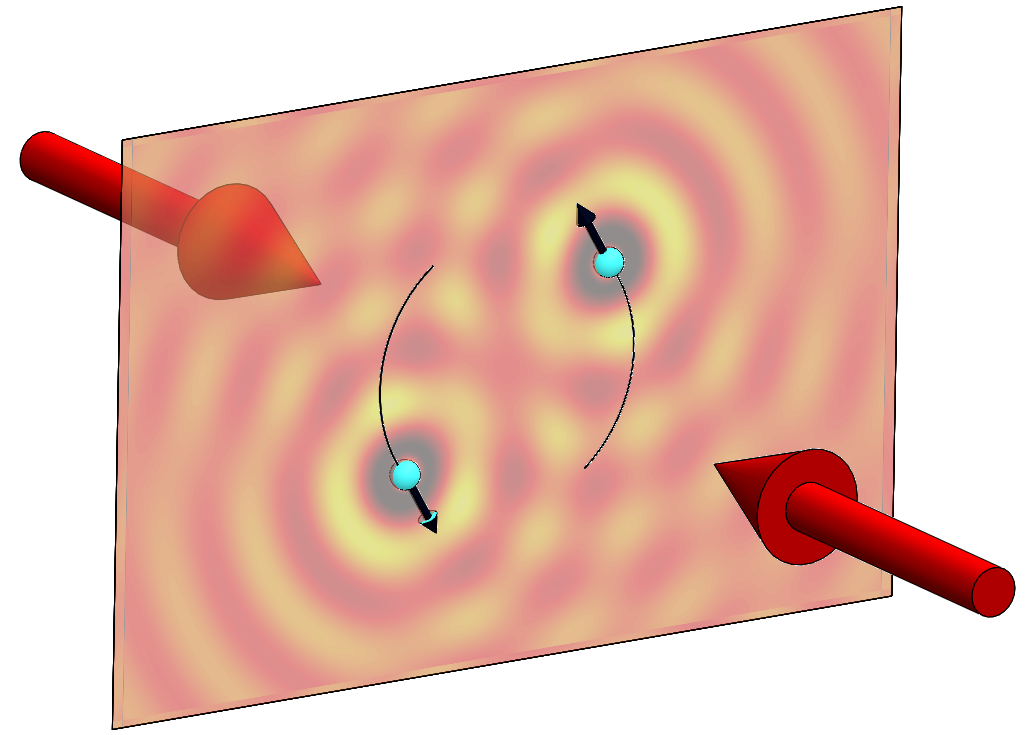}
	\end{center}
	\caption{\label{scheme}Two atoms evolve in the $z = 0$ plane, trapped in 2D by counter-propagating plane-waves, with wave vector orthogonal to that plane. The light exchange between the atoms induces the 2D two-body optically coupled dynamics.}
\end{figure}

In this work, we consider a pair of cold atoms to demonstrate optically bound motion in the absence of non-radiative friction. As the atoms are confined in two dimensions by two counterpropagating lasers (see Fig.\ref{scheme}), the angular momentum becomes a conserved quantity, which guarantees an everlasting motion. We also study the response of the two-body system to the detuning between the pump and the atomic transition, a parameter which has not been considered previously, for dielectric spheres. 

After introducing the dipole model in section II, we study in section III the scaling laws for bound states of pairs of atoms without angular momentum, confronting our findings to known results on dielectric spheres~\cite{golovchenko1989,golovchenko1990}. We then turn to the more general situation in section IV, setting a finite angular momentum for the pair of atoms, and discuss the new regimes reached thanks to dynamical stability. Finally, we draw our conclusions and discuss possible future works.

\section{Model}

We consider a system composed of two two-level atoms of equals masses $m$ which interacts with the radiation field. Using the dipole approximation, the atom-light interaction is described by
\begin{equation}
H_{AL}=-\sum_{j=1}^2 \mathbf{D}_j\cdot\mathbf{E}\left(\mathbf{r}_j\right), \label{H}
\end{equation}
where $\mathbf{D}_{j}$ is the dipole operator and $\mathbf{E}\left(\mathbf{r}_j\right)$ the electric field calculated at the center-of-mass $\mathbf{r}_j$ of each atom. { Under the dipole approximation the light wavelength is much larger than the typical atom size, which motivates a classical description of the center-of-mass motion. Considering the linear optics regime of low pump intensities, the Heisenberg equations of motion for the expected values of the atomic dipoles $\beta_j$ and of their center-of-mass positions reduce to the following classical dynamical system~\cite{kaiser2012}:
\begin{eqnarray}
\dot{\beta}_{j}&=&\left(i\Delta-\frac{\Gamma}{2}\right)\beta_{j}-i\Omega -\frac{\Gamma}{2}G\left(\left|\mathbf{r}_{j}-\mathbf{r}_{l}\right| \right)\beta_{l},\label{b1}
\\ \ddot{\mathbf{r}}_{j}&=& -\frac{\hbar\Gamma}{m}\mbox{Im}\left[\nabla_{\mathbf{r}_{j}}G\left(\left|\mathbf{r}_{j}-\mathbf{r}_{l}\right| \right)\beta_{j}^{*}\beta_{l}\right], \quad l\neq j,\label{r1}
\end{eqnarray}
The present work assumes that the atoms are trapped in the $z=0$ plane, a situation which can be obtained with a pair of counter-propagating plane waves, for example. The strength of the atom-laser coupling in that plane is given by the Rabi frequency $\Omega$ of the resulting stationary wave, whose frequency $\omega_L=c k$ is tuned close to the two-level transition frequency $\omega_{at}$. $m$ is the mass of the atom, $\Gamma$ is the decay rate of its excited state and $\Delta=\omega_{L}-\omega_{at}$ the detuning between the laser and the associated atomic transition. The light-mediated long-range interaction between the atoms is given, in the scalar light approximation, by the Green function
\begin{equation}
G\left(\left|\mathbf{r}_{j}-\mathbf{r}_{l}\right| \right)=\frac{e^{ik \left|\mathbf{r}_{j}-\mathbf{r}_{l}\right|}}{ik \left|\mathbf{r}_{j}-\mathbf{r}_{l}\right|},
\end{equation}
where $G$ has been obtained from the Markovian integration over the vacuum modes of the electromagnetic field~\cite{kaiser2012}, which makes of the two-atom system an open one. 

Eq.(\ref{b1}) thus describes the dipole dynamics under the influence of the pump and of the radiation of the other dipole, in the linear optics regime, whereas Eq.(\ref{r1}) captures the atom motion under the effect of the optical force generated by the other atom (the 2D trapping beams do not generate any force in the plane orthogonal to their propagation). Note that, while the dipole amplitude $\beta_j$ responds linearly to the light field $\Omega$, their mutual coupling with the centers-of-mass dynamics is responsible for the nonlinearity emerging in Eqs.(\ref{b1}--\ref{r1}). Similar emergent non linearities are at the origin of photon bubbles~\cite{kaiserbubbles2012} and optomechanical pattern formation~\cite{labeyrie2014}.

This coupled dipole model has been shown to provide an accurate description of many phenomena based on cooperative light scattering, such as the observation of a cooperative radiation pressure force~\cite{kaiser2010,1kaiser2010}, superradiance~\cite{michele2016} and subradiance~\cite{robin2016} in dilute atomic clouds, linewidth broadening and cooperative frequency shifts~\cite{bromley2016,1browaeys2016}.In such a description quantum matter wave effects are neglected. Also quantum optics effects, expected to arise for strong pump intensities are not taken into account. In particular Eq.~(\ref{r1}) neglects the contribution of the stochastic heating which rises from the random nature of the spontaneous emission recoils. As a first step, this hypothesis is a convenient simplification since a bound motion can have its origin only from deterministic force laws. The implementation of a more realistic stochastic dynamics, for example using a Langevin scheme, will be an important next step toward demonstrating the feasibility of bound states with cold atoms.
  
Central force problems, as the one described by Eqs.~(\ref{b1}-\ref{r1}), are best studied in the relative coordinate frame, so we define the following coordinate transformations,
\begin{eqnarray}
b&=&\beta_{1}-\beta_{2},
\\ B&=&\frac{\beta_{1}+\beta_{2}}{2},
\\ \mathbf{r}&=&\mathbf{r}_{1}-\mathbf{r}_{2},
\\ \mathbf{R}&=&\frac{\mathbf{r}_{1}+\mathbf{r}_{2}}{2}.\label{frame_s}
\end{eqnarray}
where $B$ and $b$ are the average and differential dipoles, and $\mathbf{R}$ and $\mathbf{r}$ the center-of-mass and differential positions. As the motion is restricted to two dimensions, we choose the polar coordinates parametrization $\mathbf{r}=r\left(\cos\phi,\sin\phi\right)$, which yields the following dynamical equations:
\begin{eqnarray}
 \overset{.}{b}&=&-\frac{\Gamma}{2}\left[1-\frac{\sin kr}{kr}-i\left(2\delta-\frac{\cos kr}{kr}\right)\right]b, \label{b_s}
\\ \overset{.}{B}&=&-\frac{\Gamma}{2}\left[1+\frac{\sin kr}{kr}-i\left(2\delta+\frac{\cos kr}{kr}\right)\right]B-i\Omega,\label{B_s}
\\ \ddot{r}&=&\frac{2 L^{2}}{m^{2}r^{3}}-\frac{\Gamma\hbar k}{2m}\left(\frac{\sin kr}{kr}+\frac{\cos kr}{k^{2}r^{2}}\right)\left(4\left|B\right|^{2}-\left|b\right|^{2}\right),\label{r_s}
\\ \dot{L}&=&\frac{d}{dt}\left(\frac{mr^{2}\overset{.}{\phi}}{2}\right)=0,\label{L_s}
\end{eqnarray}
where the center-of-mass dynamics has naturally decoupled from all equations. The angular momentum $L=mrv_{\perp}/2$ of the system is a conserved quantity, with $v_{\perp}$ the magnitude of the velocity perpendicular to the inter-particle separation $\mathbf{r}=r\hat{\mathbf{r}}$. This is a fundamental difference with other optical binding systems where the angular momentum is quickly driven to zero by friction.

\section{Stability analyses for $\ell=0$}

In order to study optically bound states, we first have to identify the equilibrium points of the system. We start highlighting the absence of source term in Eq.~(\ref{b_s}), so the relative dipoles coordinate $b$ vanishes on a time scale of the order of $1/\Gamma$, which is very fast compared to the atomic motion. Therefore, the atoms can be considered always synchronized and the inter-atomic dynamics $r$ couples only to the average dipole $B$. This synchronized mode has a dimensionless energy $-\cos kr/2kr$, relative to the atomic transition, and dimensionless decay rate $1+\sin kr/kr$. It becomes clear later that around the equilibrium points of the dynamics ($r\approx (n+1/2)\pi$, $n$ integer) this rate is very close to unity, so no superradiant or subradiant behaviour is expected. 

Assuming $b=0$ for all times, it is possible to write
\begin{equation}
 B=\beta_{1}=\beta_{2}\equiv\beta e^{i\chi},
\end{equation}
with $\beta \geq 0$ and $\chi \in \mathbb{R}$, which yields in the following set of reduced dynamical equations:
\begin{eqnarray}
\overset{.}{\beta}&=&-\frac{\Gamma}{2}\left(1+\frac{\sin kr}{kr}\right)\beta-\Omega\sin\chi,\label{b3_s}
\\ \overset{.}{\chi}&=&\Delta+\frac{\Gamma}{2}\frac{\cos kr}{kr}-\Omega\frac{\cos\chi}{\beta},\label{phi_s}
\\ \ddot{r}&=&\frac{2L^{2}}{m^{2}r^{3}}-\frac{2\Gamma\hbar k}{m}\left(\frac{\sin kr}{kr}+\frac{\cos kr}{k^{2}r^{2}}\right)\beta^{2}.
\label{r3_s}
\end{eqnarray}
The equilibrium points of these equations are given by choosing $\dot{\beta}=\dot{\chi}=0$, namely,
\begin{eqnarray}
\beta_0 &=& 2\frac{\Omega}{\Gamma}\left[\left(1+\frac{\sin kr}{kr}\right)^{2}+\left(2\delta+\frac{\cos kr}{kr}\right)^{2}\right]^{-\frac{1}{2}},
\label{bpoint_s}
\\ \chi_0 &=&\arctan\left(-\frac{1+\frac{\sin kr}{kr}}{2\delta+\frac{\cos kr}{kr}}\right),\label{chipoint_s}
\end{eqnarray}
which, combined with $\ddot{r}=0$, gives the equilibrium condition for the interatomic separations
\begin{equation}
F\equiv\frac{\ell^{2}}{k^{3}r^{3}}-\frac{\Omega^{2}}{\Gamma^{2}}\frac{\frac{\sin kr}{kr}+\frac{\cos kr}{k^{2}r^{2}}}{\left(1+\frac{\sin kr}{kr}\right)^{2}+\left(2\delta+\frac{\cos kr}{kr}\right)^{2}}=0.\label{condition_s}
\end{equation}
The equilibrium points are then obtained from the zeros of $F$ (see Fig.~\ref{eqpoints_s}). In particular, Fig.~\ref{eqpoints_s}(a) illustrates the dependence of $F$ on the angular momentum and on the detuning, showing that zeros of $F$ tend to disappear as we get farther from resonance or at larger angular momentum. 
Note that Eq.(\ref{condition_s}) depends on the pump strength $\Omega/\Gamma$, laser detuning $\delta\equiv\Delta/\Gamma$ and the dimensionless angular momentum $\ell\equiv Lk/2\sqrt{\hbar \Gamma m}=kr v_{\perp}/\sqrt{32}v_{Dopp}$, where $ v_{Dopp}^2=\hbar \Gamma /2m$ corresponds to the Doppler temperature of the two-level laser cooling. This equation includes both stable and unstable equilibrium points. For the particular case $\ell=0$, the equilibrium points are given by the simplified condition $\tan kr=-1/kr$ which does not depend on the light matter coupling $\Omega$ or $\Delta$, but only on the mutual distance between the atoms. The details of the light-matter interaction will however come into play when the stability of these equilibrium points is considered.
\begin{figure}[h]	
	\begin{center}
		\includegraphics[scale=0.35]{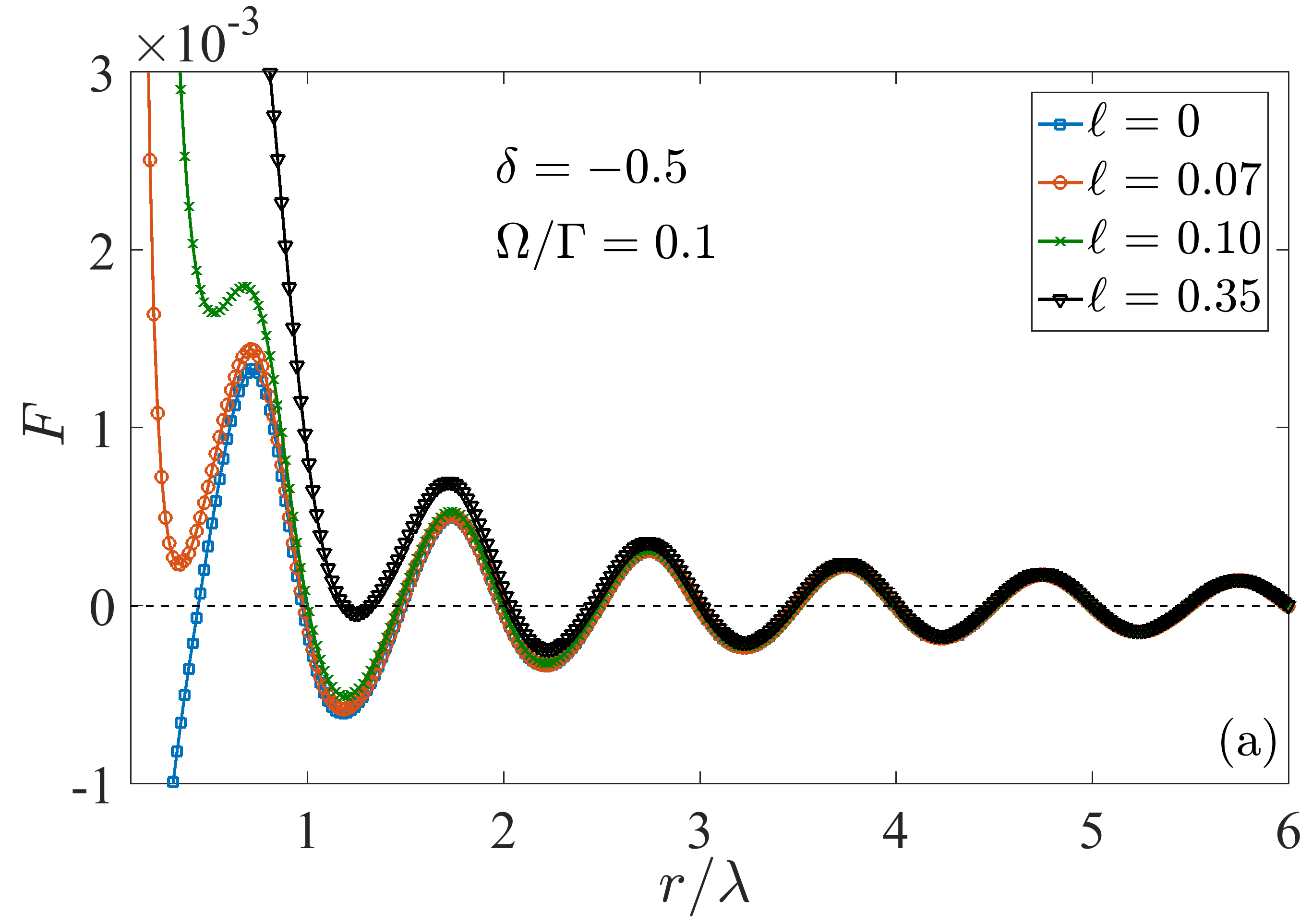}
		\includegraphics[scale=0.35]{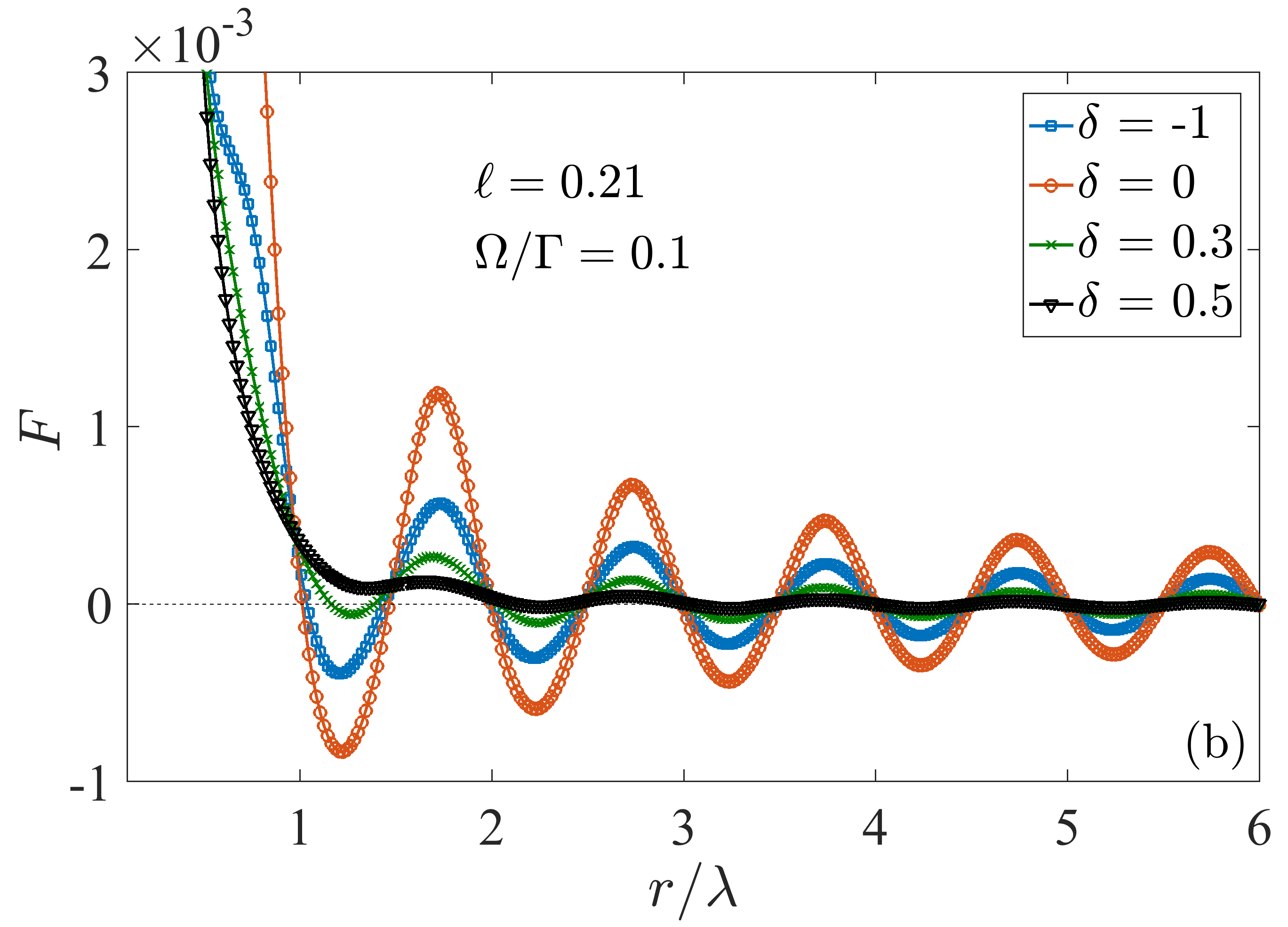}
	\end{center}
	\caption{\label{eqpoints_s}Two-atoms equilibrium distances pattern for different $\ell$ (a) and different detunings (b).}
\end{figure}

\begin{figure}[htb]	
	\begin{center}
		\includegraphics[scale=0.35]{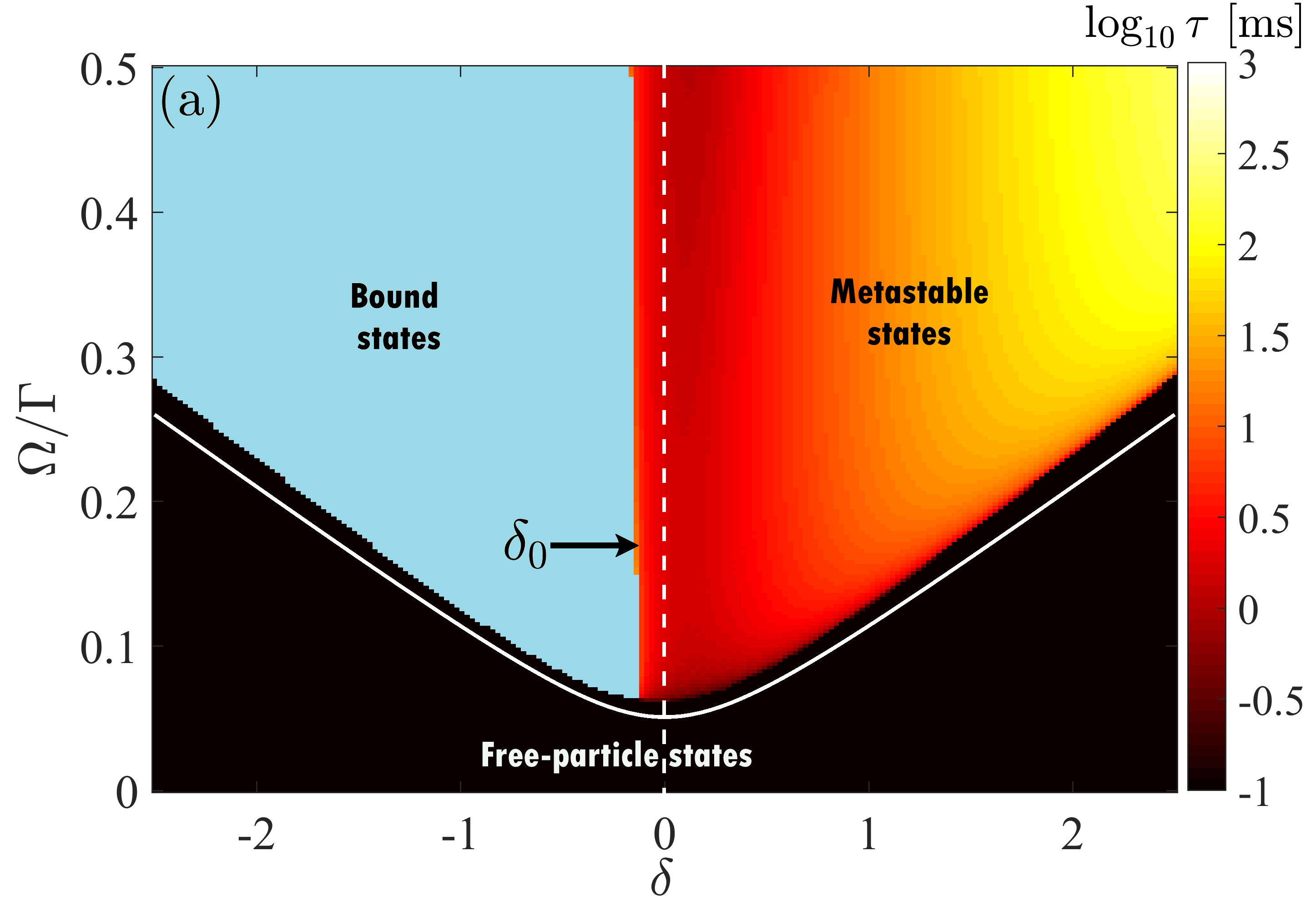}
		\includegraphics[scale=0.43]{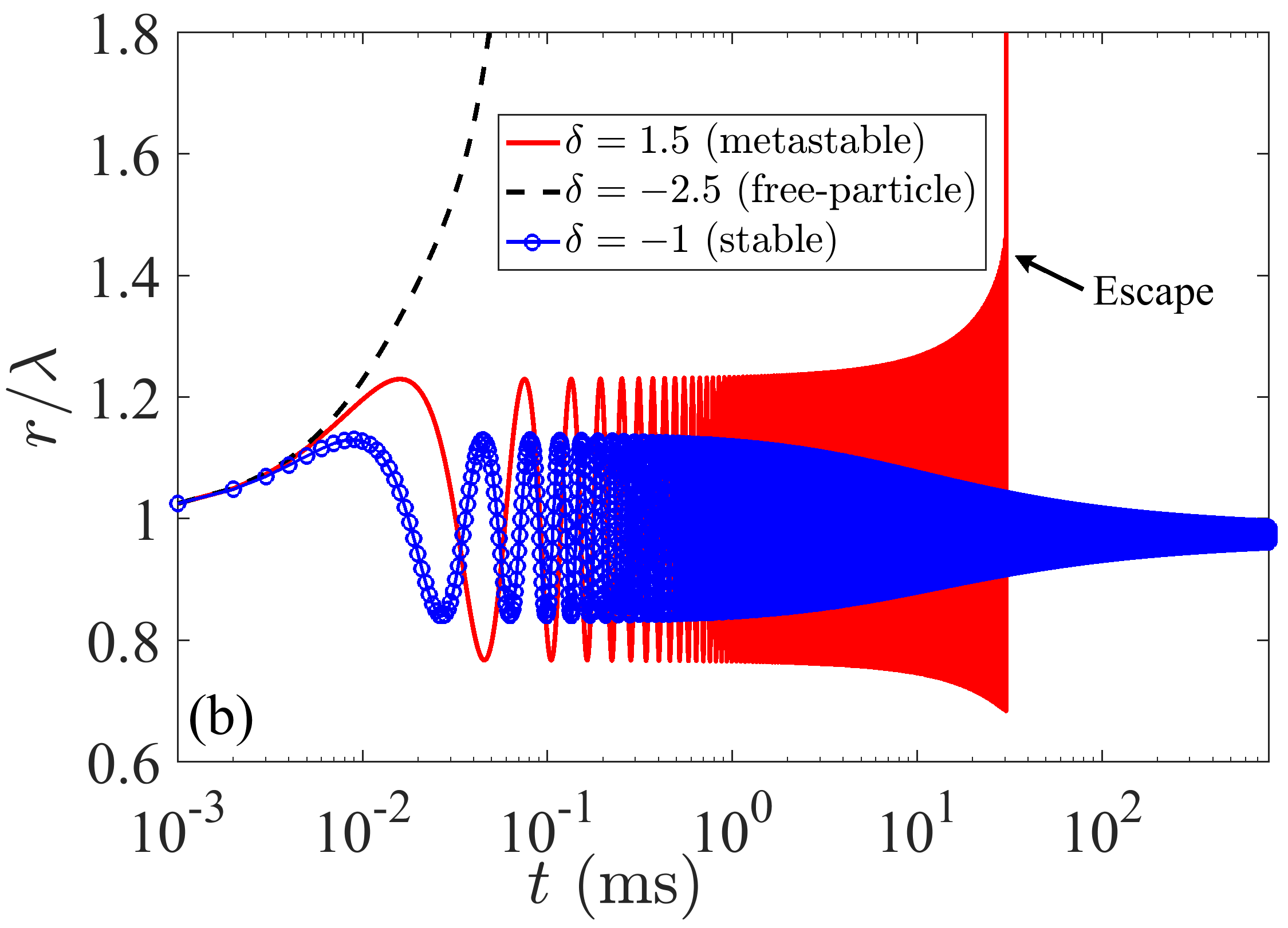}
	\end{center}
	\caption{\label{diagraml=0}(a) Stability diagram around the equilibrium point $r \approx \lambda$ for the 1D dynamics with $\ell=0$. Free-particle states are found for low laser strength (black region), bound states on the red-detuned side (light blue area) and metastable states on the blue-detuned side (color gradient). (b)  Typical dynamics of free-particle, bound and metastable states around the equilibrium point. Simulations realized with particles with an initial temperature $T=1\mu K$, as throughout this work.}	
\end{figure}

As pointed out in Ref.\citep{sheng2005}, linear stability may not be sufficient to provide a phase diagram with bound states. We thus choose to study its stability by integrating numerically the dynamics, starting with a pair of atoms with initial velocities, and moving around the equilibrium separation $r\approx\lambda$. We first focus on states without angular momentum, so the atoms are chosen with opposite radial velocities $v_\parallel$, parallel to interatomic distance. We compute the escape time $\tau$ at which the atoms start to evolve as free-particles by integrating the associated one-dimensional dynamics. 

Fig.~\ref{diagraml=0}(a) presents the stability diagram of the $r\approx\lambda$ equilibrium point, for a pair of atoms with an initial temperature of $T=1\mu$K. Throughout this work, the conversion between initial velocities and initial temperature is performed merely to give a notion about the typical atomic velocities involved. We perform this conversion by using the $^{87}$Rb atom mass $m=1.419 \times 10^{-25}$kg. We have also used in all simulations the values $\lambda=780$nm and $\Gamma\approx 6$MHz. For zero angular momentum, the initial temperature is associated to the radial degree of freedom $k_B T=mv_\parallel^2$. In the lower (black) part of the diagram (low pump strength), free-particle states are always observed after a short transient $\tau \approx 0.1$ ms, which means that the optical forces are unable to bind the atoms (see trajectory for $\delta=-2.5$ in Fig.~\ref{diagraml=0}(b)). For negative $\delta$, a free-particle to bounded motion phase transition occurs as the pumping strength $\Omega/\Gamma$ is increased. In the stable phase, the atoms mutual distance $r$ converges oscillating to $r \approx \lambda$, as displayed for $\delta=-1$ in Fig.~\ref{diagraml=0}(b). We verified that these bound states, characterized by an infinite escape time $\tau$, occur around the equilibrium points $r=n\lambda$, with $n\in \mathbb{N}$. The unstable equilibrium points are found around $(n+1/2)\lambda$ and contain only free-particle states. We stress that the damping observed in the bound state region cannot be associated to a viscous medium as in~\cite{golovchenko1989,golovchenko1990}. In our case, the cooling of the relative motion for negative detunging can be understood as Doppler cooling in a multiply scattering regime~\cite{guillot2001,cheng2009,lingxiao2010}. 

For positive detuning ($\delta>0$), we also observe a free-particle to bound states phase transition, symmetric to the negative detuning case(see Fig.~\ref{diagraml=0}(a)). However we find that these bound states appear to be only metastable, with the pair of atoms separating on large time scales (see Fig.~\ref{diagraml=0}(b)). As presented in Fig.~\ref{escape_vs_delta}, the binding time tends to grow exponentially with the detuning $\delta$ in the metastable regime.  We have verified that the oscillations of an atom in the optical potential of another atom kept at a fixed position also results in an increase (decrease) of kinetic energy for positive (negative) detuning. These results of mutual heating or cooling is thus reminiscent of the asymmetry reported in multiple-scattering based atom cooling schemes~\cite{guillot2001,cheng2009,lingxiao2010}, but differs from previous works on optical binding of dielectric spheres, which did not study the sign of the particles refractive index~\cite{pavel2010}.

\begin{figure}[htb]	
	\begin{center}
		\includegraphics[scale=0.38]{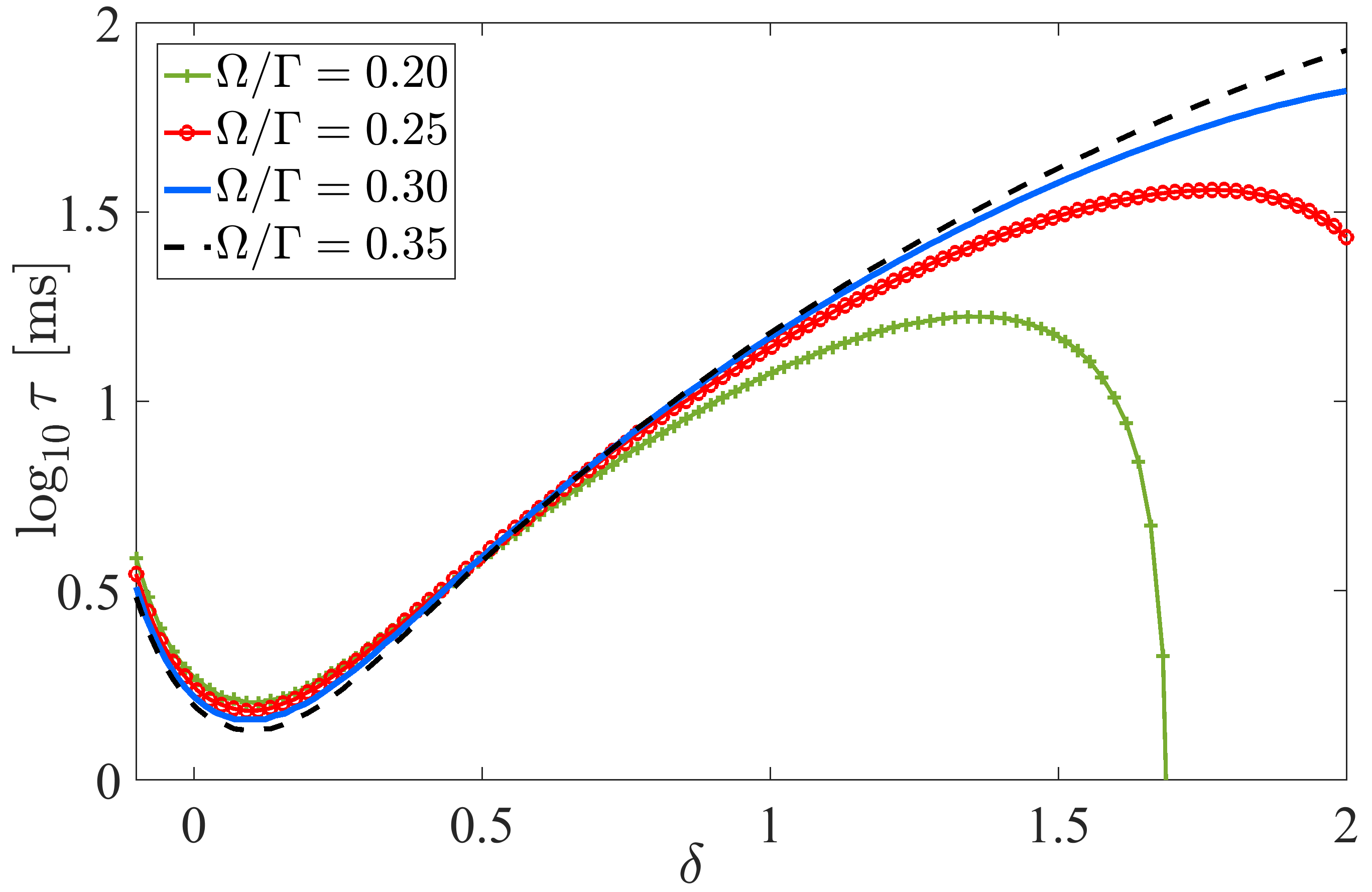}		
	\end{center}
	\caption{\label{escape_vs_delta} Logarithm of the escape time $\tau$ versus the detuning for different values of $\Omega/\Gamma$. $r \approx \lambda$ for the 1D dynamics with $\ell=0$. }	
\end{figure}
  
We have performed a systematic study of the free to bound states transition and identified the  following scaling law for the critical initial temperature $T_c$ of this transition:
\begin{equation}
\frac{T_{c}}{T_{Dopp}}=\frac{ \Omega^{2}}{\Gamma^2+4\Delta^2}\frac{16}{kr},\label{lawl=0}
\end{equation}
with $T_{Dopp}=\hbar \Gamma /2 k_B$ the Doppler temperature. This criterion can be obtained from the balance between the kinetic energy $E_\text{kin}=m k_B v_{\parallel}^2/2$ and the dipole potential induced by the interference between the incident laser beam and the scattered light field
\begin{equation}
V(r)=\frac{4\hbar\Omega^2}{\Gamma^2+4\Delta^2}\frac{\cos kr}{kr}.
\end{equation}
We recall that here we are considering zero angular motion dynamics, and the temperature is thus associated to the parallel (radial) velocity of the two atoms ($k_BT=m\langle v_\parallel^2 \rangle$). Eq.\eqref{lawl=0} is valid for atoms at large distances ($r\gg\lambda$) since for short distances, corrections due to their coupling should be included. We note that this scaling law corresponds to the law derived for dielectrics particles, where the interaction is given by $W=-\frac{1}{2}\alpha^2E^2k^2\cos kr/r$ ~\cite{golovchenko1989}, where the $\alpha^2$ scaling of the polarisability is the indication of double scattering. The corresponding scaling law for two-level atoms yields a dipole potential ($ \propto \Omega^2$) but with double scattering and a corresponding square dependence of the atomic polarisability, which at large detuning scales as $\alpha^2\propto 1/\Delta^2$. The difference to previous work lies in the fact that we do not have an external friction or viscous force, which would damp the atomic motion independently from the laser detuning. The metastable phase region is thus a novel feature for cold atoms compared to dielectrics embedded in a fluid.

A fine analysis of the transition between bound states to metastable states in Fig.~\ref{diagraml=0}(a) shows a small shift $\delta_0$ compared to the single atom resonance condition. The origin of this shift can partially be understood by the  cooperative energy shift of the two--atom state at the origin of the dynamics for these synchronized dipoles. Indeed, despite the synchronized pair of atoms does not present any superradiant or subradiant effect, it possesses a finite energy $\delta_s=\cos kr/2kr$, as can be observed in Eq.\eqref{B_s}: This shift will thus be strongest for close pairs of atoms ($kr\approx \lambda/2$). In addition to this cooperative effect, we identified an additional dependence on the velocity, so the total shift $\delta_0$ scales as
\begin{equation}
\delta_0\approx-\frac{\cos kr}{2kr}-\frac{kv_{\parallel}\left(\tau\right)}{\Gamma}.
\end{equation}

\begin{figure}[htb]	
	\begin{center}
		\includegraphics[scale=0.50]{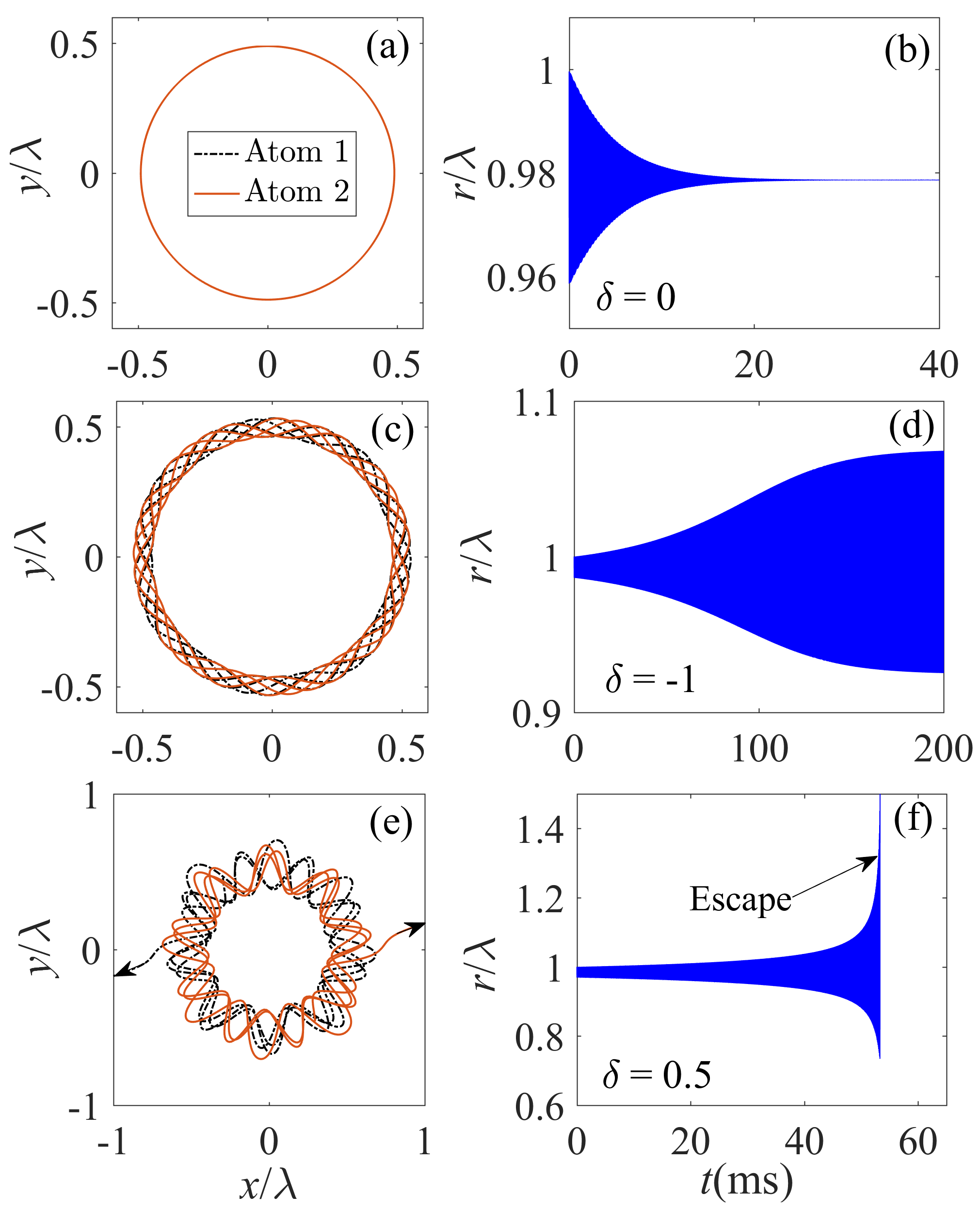}
	\end{center}
	\caption{\label{orbits} Rotating states for the pair of atoms for (a--b) Fixed inter-particle distance, (c--d) Vibrational mode and (e--f) Metastable state. As the orbits of both atoms coincides in (a), only a single orbit appears displayed. Simulations realized with $\ell=0.09$, $\Omega/\Gamma = 0.25$ and $T=1\mu K$, where the $^{87}$Rb atom mass was adopted. }
\end{figure}

\section{Stability analyses for $\ell>0$}

An additional novel feature emerging from the frictionless nature of the cold atom system is the conservation of the angular momentum during the dynamics. This leads to a striking difference to optical binding with dielectric particles as we can obtain rotating bound states. Examples of such states are shown in Fig.~\ref{orbits}: the pair of atoms can reach a rotating bound states with fixed inter-particle distance on resonance (see Figs.~\ref{orbits} (a) and (b)), but it may also support stable oscillations along the two-atom axis far from resonance (see Figs.~\ref{orbits} (c) and (d)), analogue to a molecule vibrational mode. As in the 1D case, the atoms may remain coupled for long times, before eventually separating (see Figs.~\ref{orbits} (e) and (f)). As can be seen in Eq.\eqref{condition_s}, for $\ell \neq 0$ the stationary points are circular orbits in the plane $z=0$ instead of fixed points. We note that high values of angular momentum $\ell$ strongly modifies the equilibrium point landscape, suppressing the low-$r$ equilibrium points~\cite{supplmat}. 
\begin{figure}[!h]	
	\begin{center}
		\includegraphics[scale=0.36]{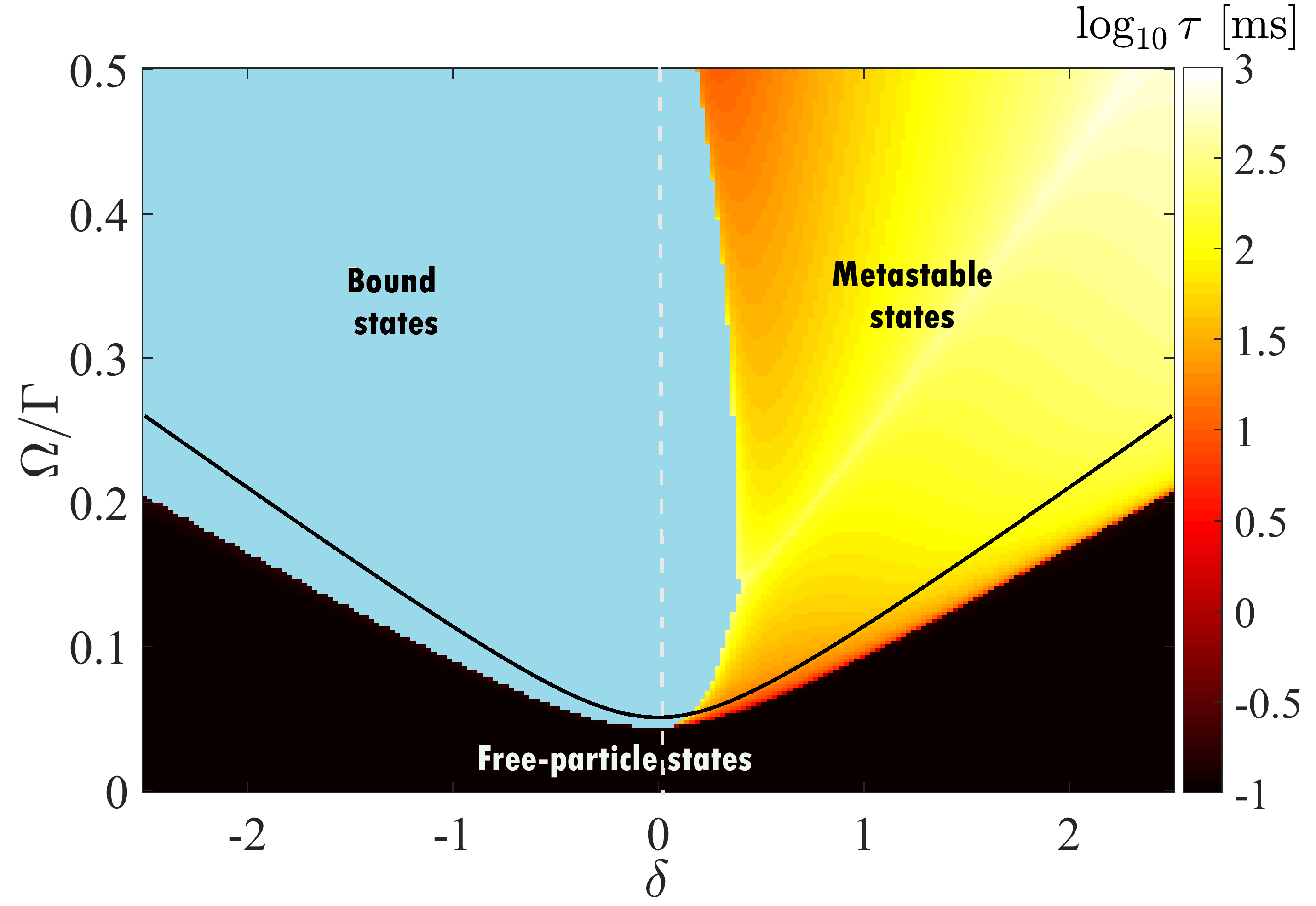}
	\end{center}
	\caption{\label{diagramlnon0} Stability diagram around the equilibrium point $r \approx \lambda$ for $\ell=0.09$. Free-particle states are found for low pump intensity (black region), bound states on the red-detuned side (light blue) and metastable states in the blue-detuned side, where the color gradient denotes the state lifetime. Simulations realized with particles with an initial temperature  $T=1\mu$K, where the conversion between temperature and velocity was realized using the Rb atom mass.}
\end{figure}

Rotating bound states are characterized by both radial $v_{\parallel}$ and tangential $v_{\perp}$ velocities, the latter being associated to the conserved angular momentum. For simplicity, we focus on initial states of atoms with purely tangential and opposite velocities: $k_B T=mv_\perp^2$, neglecting thus any initial radial velocity, although it may appear dynamically. A phase diagram for the $r \approx \lambda$ and $\ell=0.09$ (corresponding to an initial temperature $T=1\mu K$) states is presented in Fig.~(\ref{diagramlnon0}), where the escape time $\tau$ has been computed following the same procedure as before. Let us first remark that the purely-bounded to metastable transition is not delimited by a sharp transition anymore, in contrast to the $\ell=0$ case. For $\ell \neq 0$, this energy shift varies nonlinearly according to the field pump strength $\Omega/\Gamma$, allowing dynamically stable bound states for $\delta>\delta_0$, including the resonant line. The stable-metastable phase transition for  $\ell \neq 0$ covers a larger part of the phase diagram, so the introduction of an angular momentum in the system allows to reach bound states for ranges of parameters where they do not exist at $\ell=0$.

An estimation of the stability criterion for rotating bound states can be obtained via the following analysis: We consider rotating states with large inter--particle distance ($r\gg \lambda$) where the initial kinetic energy is associated only to the rotational degree of freedom: $k_BT=m\langle v_\perp^2\rangle$ and $v_\parallel=0$. The conservation of the angular momentum $L=mrv_\perp/2$ implies that over a displacement of $\delta r=\lambda/2$ necessary to escape the radial potential well $\delta V=-4\Gamma\hbar\Omega^{2}/3\pi(\Gamma^{2}+4\Delta^{2})$, only a portion of the kinetic energy $\delta E=mv_\perp L^2\delta r/2r$ is transferred to the radial degree of freedom. This leads to the stability transition law:
\begin{equation}
\frac{T_{c}}{T_{Dopp}}=\frac{8}{\pi}\frac{\Omega^{2}}{\Gamma^2+4\Delta^2},  \label{lawlnon0}
\end{equation}
A systematic numerical study of the stability of rotating bound states, tuning their angular momentum, initial inter--particle distance and temperature, shows an excellent agreement with Eq.\eqref{lawlnon0}.

Let us first comment that the diagram of the $\ell \neq 0$ case presents lifetimes for the metastable states which are much longer than for the $\ell=0$, see Fig.\ref{diagramlnon0}. Moreover, Eq.\eqref{lawlnon0} differs substantially from the $\ell=0$ case in that the distance between the atoms has disappeared: In other words, for a given angular momentum $l$, rotating pairs of bound states will be stable below a certain kinetic energy which does not depend of their distance. Thus, the presence of a conserved angular momentum strongly promotes the stability of the system, and is particular promising for the optical stabilization of macroscopic clouds. This dynamical stabilization of optical binding in frictionless media is a important novel feature since it opens the possibility of enhanced long range and collective effects in cold atoms and beyond.

\section{Conclusions}

In conclusion, our study of optical binding in cold atoms has allowed to recover the prediction of optical binding of dielectric particles for negative detuning, where the viscosity of the embedding medium is replaced by a diffusive Doppler cooling analogue. For positive detuning, we find a metastable region, as Doppler heating eventually leads to an escape of the atoms from the mutually induced dipole potential. We also identified a novel dynamical stabilization with rotating bound states. We have shown that pairs of cold atoms can exhibit optically bound states in vacuum. The absence of non-radiative damping in the motion allows for a new class of dynamically bound states -- a phenomenon not present in other optical binding setups. While this demonstration of optical binding for a pair of particles paves the way for the study of this phenomenon on larger atomic systems, the generalization of these peculiar stability properties will be an important issue to understand the all-optical stability of large clouds.

One interesting generalization is the study of optical forces in astrophysical situations. Whereas radiation pressure forces are well studied and participate for instance in the determination of the size of a star, dipole forces are often neglected~\cite{ritsch2013}. The possibility of trapping a large assembly of particles in space would allow to consider novel approaches in astrophysical imaging~\cite{labeyrie1999, fournier2014} and could shed additional light on the motion of atoms such as the abondant hydrogen around high intensity regions of galaxies, where even small corrections to the pure gravitational attraction might be important~\cite{sanders1991}.

In the context of cold atoms scattering light, an important issue is the
one of stochastic processes due to spontaneous emission. In order to obtain
more complete predictions on the possibility to optically bind atoms, one
may for example implement a Langevin dynamics to study the effect of
stochastic processes on the stability diagram. This may also be relevant to
account for collective diffusion or collisions as one consider many-atom
systems.

\acknowledgments
C.E.M. and R.B. hold Grants from São Paulo Research Foundation
(FAPESP) (Grant Nos. 2014/19459-6, 2016/14324-0, 2015/50422-4  and 2014/01491-0). R.B. and R.K. received support from project CAPES-COFECUB (Ph879-17/CAPES 88887.130197/2017-01). We thank fruitful discussions with N. Piovella.

\bibliographystyle{apsrev4-1}
\bibliography{ref}


\appendix 

\section{}
In Fig.~\ref{r=2_s} the phase diagram of the $r\approx2\lambda$ equilibrium point is shown: The pair of atoms presents a higher threshold in pump strength to become stable, in order to compensate for their weaker interaction at a larger distance.
\begin{figure}[h]	
	\begin{center}
		\includegraphics[scale=0.37]{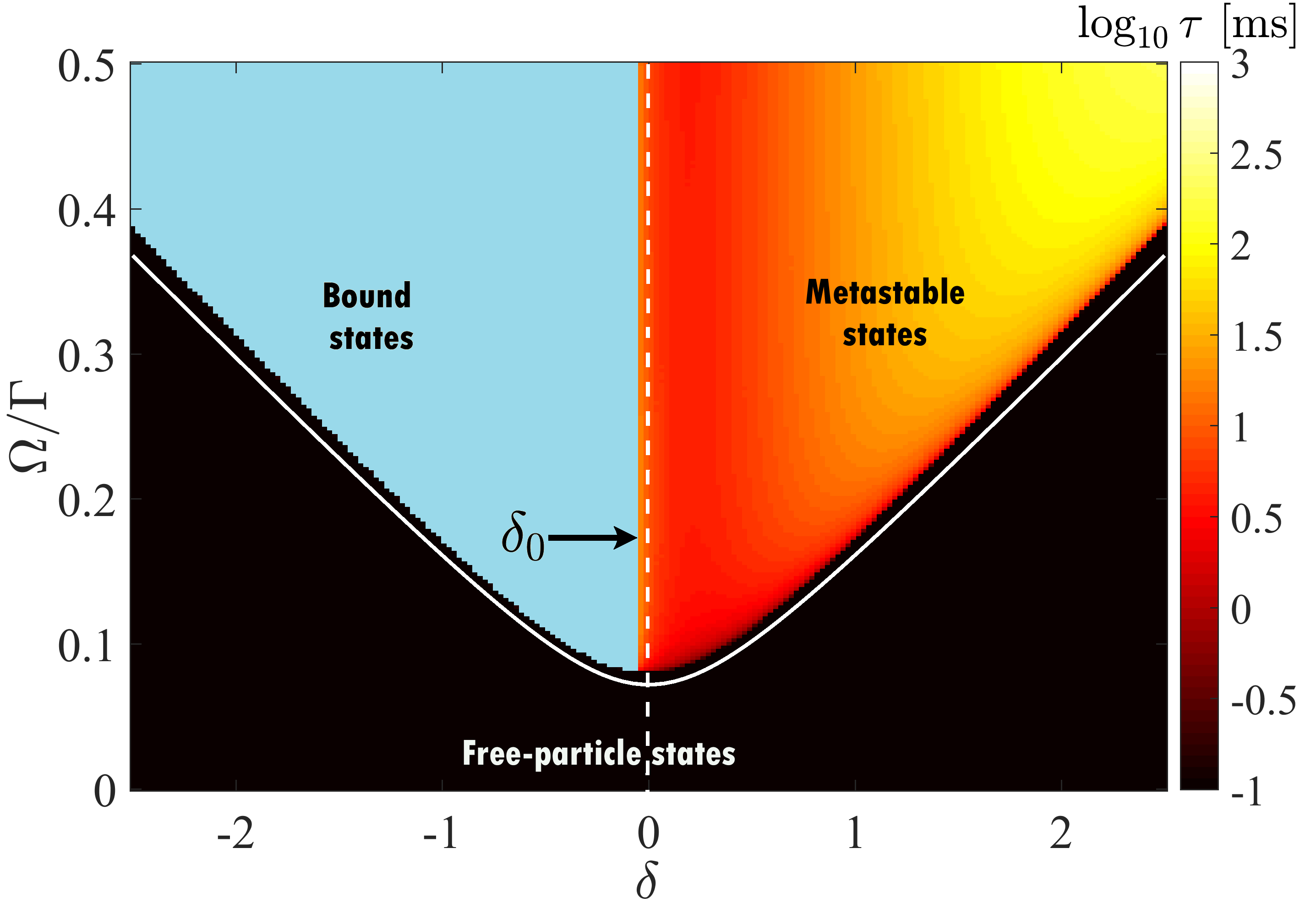}
	\end{center}
	\caption{\label{r=2_s}Stability diagram for the equilibrium point around $r \approx 2\lambda$.}
\end{figure}

\end{document}